# LOCO: A Low-Cost SNU-Self-Resilient Latch Using an Output-Split C-Element

Ruijun Ma[1,5] · Xin Chen[1] · Xiaoqing Wen[2] · Hui Xu[1] · Shengnan Ye[3] · Chuanjian Zhang[1] · Senling Wang[4]

**Abstract**

As the CMOS technology enters nanometer scales, integrated circuits (ICs) become increasingly sensitive to radiation-induced soft errors, which can corrupt the state of storage elements and cause severe reliability issues. Many hardened designs have been proposed to mitigate soft errors by using filtering elements. However, existing filtering elements only protect their inputs against soft errors and leave their outputs unprotected. Therefore, additional filtering elements must be added to protect outputs, resulting in extra overhead. In this paper, we first propose a novel Output-Split C-element (OSC) to protect both its input and output nodes, and then a novel LOw-COst single-node-upset (SNU) self-resilient latch (LOCO) to use OSCs to achieve both soft error resilience and low overhead. The usage of OSCs effectively reduce the short-circuit current of the LOCO latch during switching activities. Furthermore, the usage of clock gating and high-speed path reduces power consumption and delay, respectively. Compared with state-of-the-art SNU-resilient hardened designs, the LOCO latch achieves 19% fewer transistors, 63.58% lower power, 74% less delay, and 92% lower power-delay-product (PDP) on average. In addition, the LOCO latch exhibits better stability under variations in PVT (Process, Voltage, and Temperature).

## 1 Introduction

Safety-critical systems, such as autonomous cars, require not only high performance but also high reliability. Advanced technology nodes with lower supply voltages and smaller capacitances can achieve high performance. However, reduced supply voltages and capacitances make modern circuits more and more sensitive to radiation-induced soft errors, which often lead to severe reliability problems [1-4].

Soft errors are mainly caused by the impact of radiation particles, such as protons, neutrons, and heavy ions on the sensitive regions of a device [2]. The collision between the radiation particle and silicon atoms in the sensitive region can generate excessive electron-hole pairs, which can be collected by the drain of a transistor and cause a transient current pulse [3]. This transient current pulse may disturb the functional working of ICs and pose a severe threat to the reliability of safety-critical systems [6].

If a particle strikes a storage element (a latch or a flip-flop) within a hold mode, its state can be changed, which is called Single-Event-Upset (SEU) [4]. When a single event only affects a single node of a storage element at a time, the flip of this single node is defined as a Single-Node-Upset (SNU) [5]. The radiation particles reach devices on the ground are secondary particles, which are less energetic than primary particles (released from solar wind and galactic cosmic rays) [7]. The striking of these low energy particles on a device on the ground usually induce SNUs [5-6]. Therefore, this paper addresses the radiation-induced SNUs in a latch.

Radiation Hardened by Design (RHBD) effectively protects storage elements from soft errors, and many hardened latch designs have been presented [10-17]. STAHL [10], and HLR [11] are soft error tolerant hardened latch designs, which use dual-modular redundancy and cannot restore corrupted logic values. HiPeR [12] and Ref. [17] are partially self-resilient hardened latches, which partly sacrifice their soft error tolerability to achieve high performance. Fully self-resilient hardened latch designs, such as RFC [13] and ISEHL [14], can restore all their internal nodes and achieve high reliability.

The above-mentioned hardened latch designs rely on filtering elements (e.g., C-element, Schmitt-Trigger (ST) [8]) to block input soft errors from propagating to outputs. However, existing filtering elements suffer from a critical vulnerability: ***their outputs remain unprotected***, that is, an erroneous value at the output can propagate to downstream gates. If this erroneous value feeds back to an input of the same filtering element, it can be latched. To prevent this condition, additional filtering elements are typically placed after these vulnerable outputs. Therefore, a filtering element with the capability to protect both its input and output nodes is a desired design, which can help a latch design to achieve a good trade-off between high reliability and low cost.

This paper has two major contributions: (1) A novel Output-Split C-element (OSC), which is the first filtering element that can protect both its input and output nodes. (2) A novel Low-Cost SNU self-resilient hardened latch design (LOCO), which is the first to use OSCs to achieve full soft error resilience and low overhead. The usage of OSCs can effectively reduce the short-circuit current of the proposed LOCO latch as well as the short-circuit power [9]. In addition, a high-speed transmission path is used to reduce the propagation delay and clock gating is used to reduce power consumption. Simulation results show that the proposed LOCO latch has better performance in terms of power, delay, and area overhead while maintaining fully SNU-immunity and self-resilience. Process, Voltage, and Temperature (PVT) and Monte Carlo (MC) simulation experiments prove the stability of the proposed LOCO latch under PVT variations.

The rest of the paper is organized as follows: Section 2 introduces existing filtering elements and hardened designs. Section 3 presents the proposed OSC and the LOCO latch, including their structures and operation details. Section 4 shows evaluation results, including reliability comparison, overhead comparison, relative overhead comparison, short-circuit current analysis, PVT simulations, and Monte Carlo simulations. Section 5 concludes this paper.

## 2 Typical Hardened Latch Designs

Fig. 1 shows various existing filtering elements, which are the dual-input inverter, the clock-controlled dual-input inverter, the C-Element (CE), the clock-controlled C-Element, the Schmitt-Trigger (ST) [8], and the Scan-Enable (SE)-signal-controlled C-Element from Fig. 1(a) to Fig. 1 (f). All of these elements suffer from a common problem that their outputs remain unprotected. Unfortunately, state-of-the-art hardened latch designs are commonly based on these filtering elements.

**a. STAHL**

Fig. 2(a) shows STAHL [10], which is the first hardened latch design with high defect detectability by using the design-for-test (DFT) technique. Added DFT structures cost more area overhead and power. STAHL also has large propagation delay due to its long propagation path. Furthermore, STAHL is a soft error tolerant design, since it does not use additional filtering elements to protect its feedback loops and its internal nodes cannot be resilient to SNUs. Therefore, its reliability is compromised.

**b. HLR**

Fig. 2(b) shows HLR [11], which employs two feedback loops and a C-element to mitigate SNUs. However, like STAHL, it lacks additional filtering elements to protect these feedback loops. Consequently, its internal nodes are not self-resilient to SNUs and its robustness against SNUs remains compromised. HLR sacrifices its robustness against SNUs to achieve low overhead.

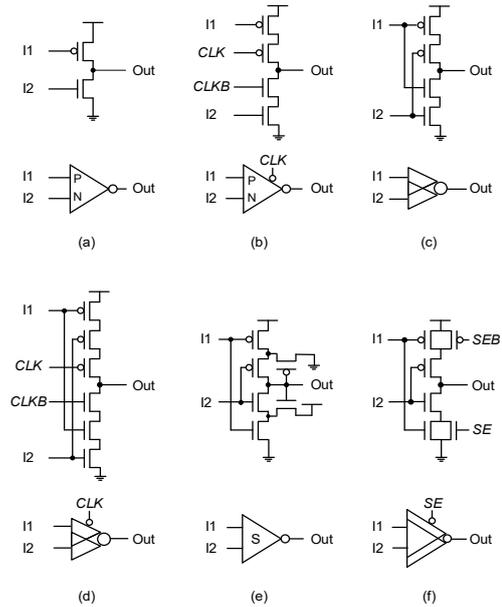

**Fig. 1** Commonly used filtering elements. **a** Dual-input inverter, **b** Clock-controlled dual-input inverter, **c** C-Element (CE), **d** Clock-controlled C-Element, **e** Schmitt-Trigger (ST), and **f** SE-signal-controlled C-Element

**c. RFC**

Fig. 2(c) shows the RFC latch [13] with the capability of self-resilience from any SNUs. RFC uses three mutually locked C-elements (CEs), where the two inputs of each CE depend on the outputs of the other two CEs. If an SNU occurs at the output of a C-element, the resulting wrong logic value can propagate to downstream gates. Therefore, it requires additional CEs to block the propagation of the wrong logic value and achieve self-resilience. The use of multiple CEs cost more power consumption.

**d. RFEL**

Fig. 2(d) shows the RFEL latch [15], which consists of an input-splitting Schmitt-Trigger (ST), two CEs and two inverters, facilitating mutually feeding back structures to ensure complete SNUs-resilient capability. The input-splitting Schmitt-Trigger can delay the propagation of a wrong logic value. Two CEs can stop the propagation of this wrong logic value. Therefore, the corrupted logic value can be restored. However, the ST significantly increases its delay and power overheads.

**e. Ref. [17]**

Fig. 2(e) shows a partial self-resilient latch design described in [17], comprising three dual-input inverters and two clock-controlled CEs. It has low area overhead and low delay due to the use of dual-input inverter and high-speed transmission path. However, the use of dual-input inverters causes SNUs occurring at Q (Qa) cannot be restored. A SNU at Q (Qa) may cause the outputs of two dual-input inverters (the upper one and the lower one) in floating states and they may change to wrong logic values simultaneously, which causes two CEs both in high-impedance state and cannot correct the SNU at Q (Qa).

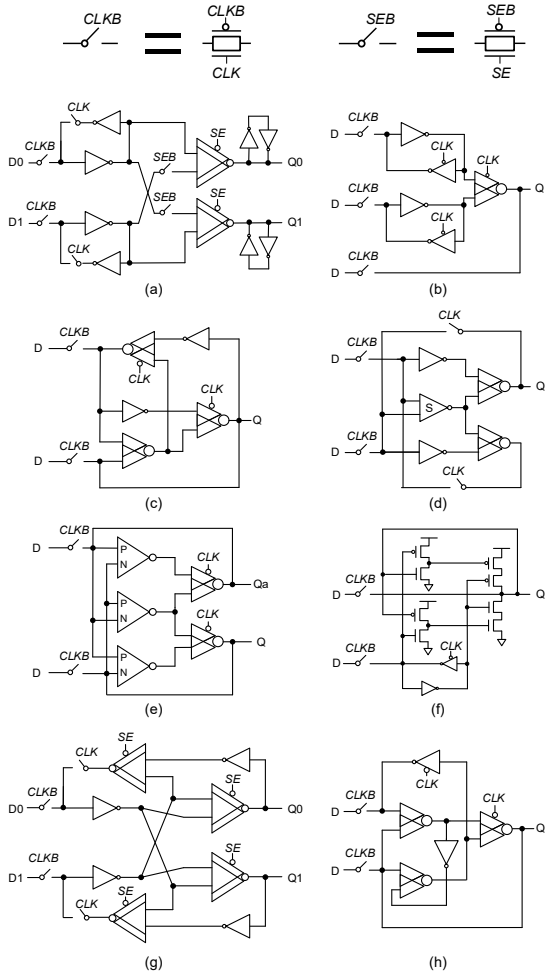

**Fig. 2** Schematics of existing hardened latches. **a** STAHL [10], **b** HLR [11], **c** RFC [13], **d** RFEL [15], **e** Ref. [17], **f** HiPeR [12], **g** HIDER [16], and **h** ISEHL [14]

### f. HiPeR

Fig. 2(f) shows the HiPeR latch [12], which is a partial self-resilient design with two feedback loops. A feedback loop is composed of two dual-input inverters and a C-element, which can restore SNUs occurring in this feedback loop to its original state. In this feedback loop, a D to Q path is used to reduce the propagation delay. The other feedback loop is composed of an inverter and a clock-controlled inverter, which has no filtering elements to block SNUs. Therefore, this feedback loop is not self-resilient to SNUs.

### g. HIDER

Fig. 2(g) shows the HIDER latch [16], which is a dual-redundancy SNU-resilient hardened latch design. It has four SE-signal-controlled C-Elements and each uses the DFT technique to improve its testability. However, these added DFT structures costs more power consumption. The long propagation path increases its propagation delay. Furthermore, these SE-signal-controlled C-Elements cannot protect their outputs. Therefore, each output requires an additional SE-signal-controlled C-Element to block SNUs from generating false feedback. Therefore, the overhead of HIDER is relatively higher.

### h. ISEHL

Fig. 2(h) shows the ISEHL latch [14], which is a self-resilient design. It comprises two transmission gates, three CEs, and two inverters. As the RFC latch, the inputs of each CE are determined by the outputs of other two CEs, and the output of each CE is connected to an additional CE to prevent the further propagation of SNUs. However, ISEHL costs more power consumption than RFC. In summary, these existing hardened designs struggle to achieve a good balance between high robustness and low cost. However, existing filtering elements cannot fill in this void. Therefore, there is an urgent need for a novel filtering element that can achieve both high robustness and low cost simultaneously.

## 3 Proposed OSC and LOCO

### 3.1 Circuit Structure and Working Details

An Output-Split C-element (OSC) is a novel filtering element that can help a hardened design to achieve both high robustness and low cost simultaneously. A novel high robustness and Low-Cost (LOCO) hardened design is based on OSCs.

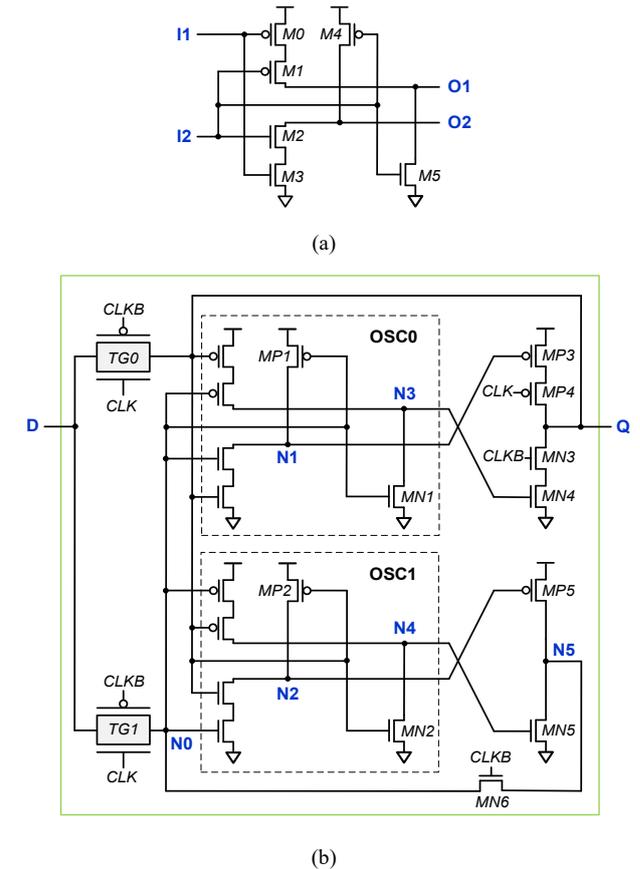

**Fig. 3 a** Schematic of the proposed Output-Split C-element (OSC), and **b** Schematic of the proposed LOCO latch

Fig. 3(a) shows the proposed OSC element, which consists of a six-transistor structure with two inputs (I1 and I2) and two outputs (O1 and O2). Input I1 drives two transistors (M0 and M3) while input I2 drives the other four transistors (M1, M2, M4, and M5). Transistors M4 and M5 are added to ensure two outputs remain identical and to restore them. For example, when inputs I1 and I2 are both at logic 0, transistors M0, M1, and M4 are ON, while M2, M3, and M5 are OFF. The outputs O1 and O2 are both initialized to logic 1. If O2 is affected by an SNU, the M4 transistor can restore the corrupted value at the output O2 to logic 1. Similarly, the M5 transistor can protect the output O1 as well.

**Table 1** The truth table of an Output-Split C-element

| I1 | I2 | O1 | O2 |
|---|---|---|---|
| 0 | 0 | 1 | 1 |
| 1 | 1 | 0 | 0 |
| 0 | 1 | 0 | Z |
| 1 | 0 | Z | 1 |

Table 1 shows the truth table of an OSC. When the two inputs are identical, the OSC operates as an inverter. When the two inputs are different, one output enters a high-impedance state, thereby retaining its previous value, while the other output becomes the inverse of input I2. Unlike a conventional C-element, an OSC has two outputs, which hold identical logic values but are physically separate. When one output is corrupted, the other remains correct. Transistors M4 and M5 are then used to correct the corrupted output. If input I1 is corrupted, the two outputs remain correct. If input I2 is corrupted, one output will be corrupted. An additional dual-input inverter can be used to block the corrupted logic value. Therefore, the OSC design hardens both its inputs and outputs against SNUs.

Fig. 3(b) shows the proposed LOCO latch, which is composed of two OSCs (OSC0 and OSC1), two transmission gates (TG0 and TG1), a dual-input inverter, and a clock-controlled dual-input inverter. D and Q stand for input and output, respectively. CLK and CLKB are clock signal and its inverse signal, respectively. A clock-controlled NMOS transistor (MN6) is placed between the node N0 and the node N5 to avoid possible current competition. Two OSCs mutually provide the capability of fully self-resilient, which greatly reduce its power consumption. Fig. 4 shows the layout of the proposed LOCO latch (Length: 17.465um and Width: 8.135um).

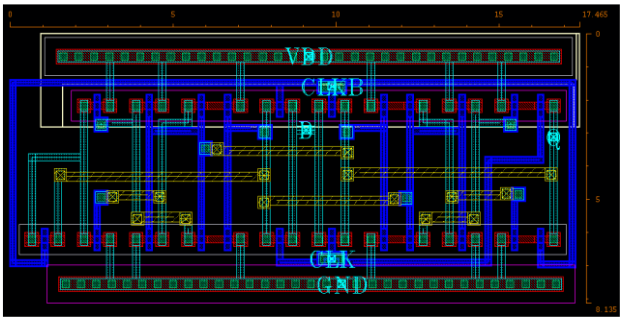

**Fig. 4** The layout of the LOCO latch (L: 17.465um and W: 8.135um).

When CLK = 1 and CLKB = 0, the proposed LOCO latch is in transparent mode. Transmission gates TG0 and TG1 are ON. Input data at D passes through TG0 to output Q. The clock-controlled transistors (MP4, MN3, and MN6) are OFF to prevent possible current competition. Consider a scenario where D = 0, nodes Q and N0 simultaneously drive OSC0 and OSC1 to deliver values to nodes N3 and N4, while MP1 and MP2 are ON to initialize nodes N1 and N2. N4 can turn ON transistor MN5 to pull down the node N5 to logic value 0. As shown in Fig. 5(a), all nodes can be initialized during a transparent mode. Black lines denote transistor-ON and gray lines denote transistor-OFF.

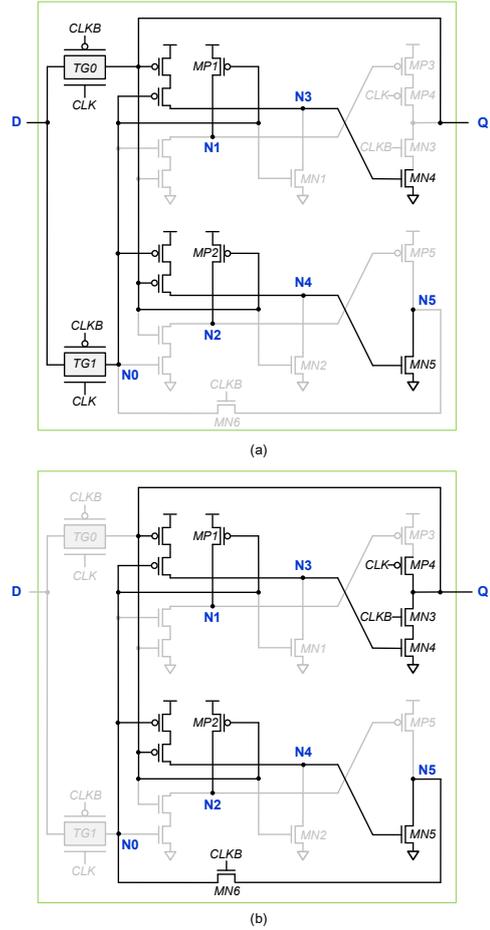

**Fig. 5** Transparent mode and hold mode of the proposed LOCO latch. **a** D = 0 in transparent mode, and **b** D = 0 in hold mode

When CLK = 0 and CLKB = 1, the proposed LOCO latch is in hold mode. The transmission gates TG0 and TG1 are OFF. The clock-controlled transistors (MP4, MN3, and MN6) are ON, and two feedback loops are formed. Fig. 5(b) shows the conducting states of transistors in hold mode while storing logic value 0.

### 3.2 Soft Error Tolerance Verification

The soft error tolerance of the proposed LOCO latch is based on two OSCs. As discussed previously, one OSC combined with a dual-input inverter can protect all its inputs and outputs from SNUs. An additional pair of one OSC and a dual-input inverter can mutually provide soft error recoverability to all internal sensitive nodes. These sensitive nodes can be divided into two cases (Case-A and Case-B). Case-A includes the inputs of two OSCs (nodes N0 and Q) and the related node N5, while Case-B includes the outputs of two OSCs (nodes N1, N2, N3, and N4).

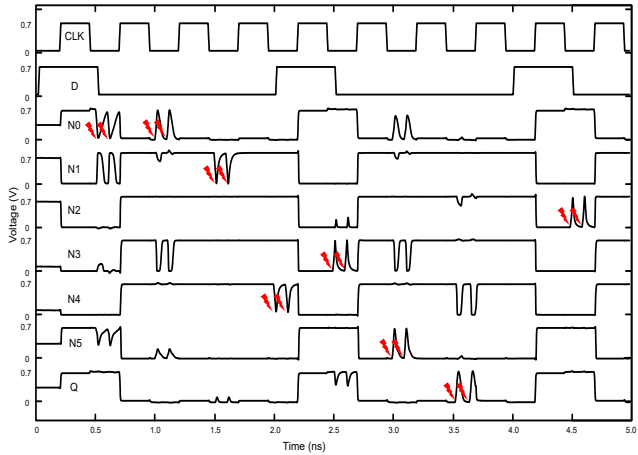

**Fig. 6** Simulation waveforms for SNU injection cases

**Case-A**: Suppose the node N0, which is at a logic 0 state, is affected by an SNU that temporally changes its value to logic 1. This erroneous value at N0 can temporally turn ON transistor MN1, and turn OFF transistor MP1. Consequently, the ON-state of MN1 pulls down node N3 to an incorrect logic 0 while node N1 remains correct. The wrong value at N3 turn OFF transistor MN4 and the correct value at N1 turn OFF transistor MP3, which corporately keep node Q at high-impedance state. Meanwhile, node N4 remains correct and continues to drive transistor MN5 to pull down node N5. Correct logic value at the node N5 propagates through the ON-state transistor MN6 to continuously restore the node N0. The corrected node N0 can then restore node N3. A similar correction analysis applies to nodes N5 and Q.

**Case-B**: Suppose the node N1, which is at a logic 1 state, is affected by an SNU that temporally changes to logic 0 and can temporally turn ON transistor MP3. At the same time, node N3 remains correct and turns ON transistor MN4, which cause node Q to be connected with VDD and GND simultaneously. However, node N0 remains correct and can drive transistor MP1 to restore node N1 to correct logic value 1. The correct logic value 1 at the node N1 turns OFF the transistor MP3 to eliminate a short-circuit condition between VDD and GND. Therefore, the output Q remains a stable state eventually. The correction analysis of N2, N3, and N4 is similar to that of N1.

A double exponential current pulse model [18-19] shown in Eq. (1) was used to simulate the SNUs injection cases, which was simulated by using SPICE with a 22nm PTM (Predictive Technology Model) process [25].

$$I_{inj}(t) = \frac{Q_{inj}}{\tau_1 - \tau_2}(e^{\frac{-t}{\tau_1}} - e^{\frac{-t}{\tau_2}}) \quad (1)$$

where $I_{inj}$ is the injected current pulse, $Q_{inj}$ is the totally collected charge, $\tau_1$ is the aggregation time constant of a particle hitting, and $\tau_2$ is the particle trajectory establishment time constant. In Eq. (1), parameters $\tau_1$ and $\tau_2$ were set to 0.1ps and 3ps, respectively, according to [20].

Fig. 6 shows the simulation results of sensitive nodes of the proposed LOCO latch injected with SNUs. The injections were set to N0 (0.50ns and 0.60ns, and 1.00ns and 1.10ns), N1 (1.50ns and 1.60ns), N2 (4.50ns and 4.60ns), N3 (2.50ns and 2.60ns), N4 (2.00ns and 2.10ns), N5 (3.00ns and 3.10ns), and Q (3.53ns and 3.65ns). The simulation results show that the proposed LOCO latch is indeed self-resilient to all potential SNU injection cases.

## 4 Experimental Results

SPICE simulation was used to evaluate all considered latch designs in the 22nm PTM with a supply voltage of 0.8 V and a clock frequency of 2GHz at room temperature. For the proposed LOCO latch, transistor aspect ratios were set to W/L = 4 for PMOS and W/L = 2 for NMOS in transmission gates TG0 and TG1. The aspect ratios of rest transistors were set to W/L = 1. All experiments were conducted under the same conditions for fair comparison.

**Table 2** Reliability and overhead comparisons among all considered latch designs

| Latch | Ref. | Full SNU Tol. | Full SNU Res. | #Trans. | Power ($\mu W$) | $Q_{cirt}$ ($fC$) | $T_{Setup}$ (ps) | $T_{Hold}$ (ps) | $T_{DQ}$ (ps) | $T_{CQ}$ (ps) | $T_{AVG.}$ (ps) | PDP ($10^{-18}J$) |
|---|---|---|---|---|---|---|---|---|---|---|---|---|
| Standard | - | × | × | 12 | 0.74 | 6.20 | 24.98 | -1.11 | 21.55 | 17.45 | 19.5 | 14.48 |
| STAHL | [10] | √ | × | 40 | 0.30 | 1.50 | 10.85 | 7.32 | 31.79 | 30.90 | 31.35 | 9.38 |
| HLR | [11] | √ | × | 24 | 0.35 | 3.10 | 11.88 | 8.63 | **5.64** | **4.38** | **5.01** | 1.74 |
| HiPeR | [12] | √ | × | 18 | 0.35 | 6.30 | 12.39 | 10.12 | 6.30 | 5.49 | 5.89 | 2.03 |
| Ref. [17] | [17] | √ | × | 22 | 0.51 | 6.80 | 12.00 | 8.34 | 6.79 | 5.60 | 6.20 | 3.13 |
| RFC | [13] | √ | √ | 24 | 0.28 | 6.80 | 12.12 | 11.28 | 6.18 | 4.98 | 5.58 | 1.59 |
| ISEHL | [14] | √ | √ | 24 | 0.67 | **11.50** | 20.52 | 12.95 | 6.14 | 4.93 | 5.54 | 3.70 |
| RFEL | [15] | √ | √ | 26 | 0.64 | 1.90 | 24.69 | 11.41 | 54.34 | 52.99 | 53.67 | 34.54 |
| HIDER | [16] | √ | √ | 40 | 0.35 | 1.7 | 16.43 | 5.92 | 25.14 | 24.18 | 24.66 | 8.57 |
| LOCO | Proposed | √ | √ | 23 | **0.18** | 6.90 | **9.06** | 10.55 | 5.86 | 4.62 | 5.24 | **0.93** |

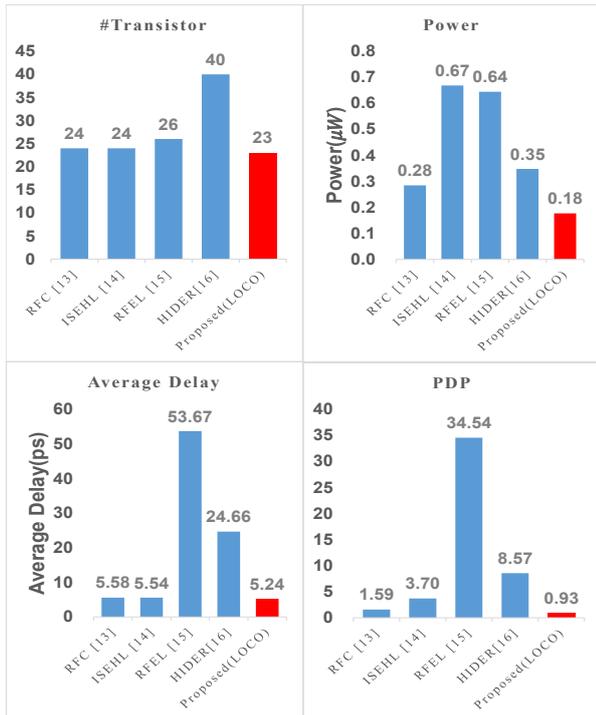

**Fig. 7** Comparison of four self-resilient hardened latch designs and the proposed LOCO latch

### 4.1 Reliability and Overhead Comparisons

Table 2 shows the comparison of reliability and overhead among all considered latch designs. The term "Full SNU Tol." denotes full SNU tolerance, and "Full SNU Res." denotes full SNU resilience. The proposed LOCO latch can tolerate soft errors and is full self-resilient to soft errors, which shows its robustness against soft errors. "#Trans" denotes the number of transistors, which is relative to its area overhead. Compared with all SNU self-resilient designs, the proposed LOCO latch has the lowest transistor count. Power consumption calculation includes both static power and dynamic power. The proposed LOCO latch has the lowest power consumption. $Q_{cirt}$ is short for the critical charge of a circuit node. The $Q_{cirt}$ is calculated as in [20]. A higher critical charge number shows a higher reliability. The critical charge of the proposed LOCO latch is only second to ISEHL [14].

Columns 8 and 9 show the setup time and hold time, respectively. The setup time of the proposed LOCO latch is the smallest, meaning that the same clock skew problem can have the least impact on it. The hold time of the proposed LOCO latch is slightly higher but still better than RFC, ISEHL, and RFEL. Columns 10, 11, and 12 show the D-to-Q delay, CLK-to-Q delay, and the average of these two delays, respectively. These delay results of the proposed LOCO latch are only second to HLR, a soft error tolerant design. The last column shows the performance comparison results, quantified by power-delay product (PDP) and calculated by Eq. (2). The PDP of the proposed LOCO latch is the best among all considered latch designs.

**Table 3** Relative overhead comparison

| Latch | Ref. | Δ#Trans. | ΔPower | Δ$T_{AVG.}$ | ΔPDP |
|---|---|---|---|---|---|
| RFC | [13] | -4.17% | -37.74% | -5.21% | -41.54% |
| ISEHL | [14] | -4.17% | -73.51% | -4.61% | -74.94% |
| RFEL | [15] | -11.54% | -72.50% | -89.22% | -97.32% |
| HIDER | [16] | -42.50% | -49.08% | -76.69% | -89.18% |
| Average | - | -19.30% | -63.58% | -74.47% | -92.34% |

$$PDP = Power * Average\ Delay \quad (2)$$

Fig. 7 compares the proposed LOCO latch with four state-of-the-art self-resilient hardened latch designs. The LOCO latch demonstrates advantages in terms of transistor count, power consumption, delay, and overall performance compared to these counterparts.

Table 3 details the relative overhead comparison between the proposed LOCO latch and these 4 designs, including their average results, calculated using Eq. (3) [24]. A positive value indicates the compared design outperforms the LOCO latch, while a negative value signifies the proposed LOCO latch's superiority. As shown in Table 3 (including the average row), all relative comparison results are negative. This confirms that the LOCO latch achieves reduced transistor count, power, and delay, alongside enhanced performance.

$$\Delta = \frac{(The\ proposed - The\ compared)}{The\ compared} \times 100\% \quad (3)$$

### 4.2 Short-Circuit Current Analysis

The proposed LOCO latch consumes significantly less power than existing hardened latch designs, primarily due to its innovative use of OSCs. Typically, dynamic power dominates total power consumption, comprising refresh transistor power (from charging/discharging) and short-circuit power [9]. RFC, ISEHL, and RFEL designs have transistor counts similar to that of the proposed LOCO latch; thus, their refresh power is likely comparable. Consequently, the substantial power reduction in the proposed design stems from differences in short-circuit power. To analyze short-circuit currents during switching in transparent mode, we measured the average current across two switching events according to Eq. (4), which integrates absolute current values from $t_0$ to $t_1$.

$$I_{(avg)} = \frac{\int_{t_0}^{t_1} |I(VDD)| dt}{t_1 - t_0} \quad (4)$$

where $I_{(avg)}$ denotes the short-circuit current, $I(VDD)$ denotes the current that flows through supply voltage, $t_0$ and $t_1$ denote the start and the end of the measurement, which is set to 50ps and 250ps, respectively.

Fig. 8 shows the average current during two switching events for the standard latch, RFC, ISEHL, RFEL, HIDER,

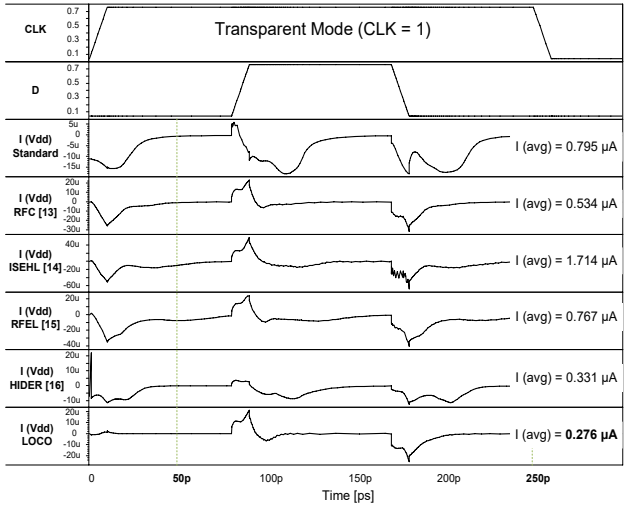

**Fig. 8** Average current of two switching activities in transparent mode

and the proposed LOCO latch in transparent mode (CLK = 1). The input D switches at 80ps and 170ps. The proposed LOCO latch exhibits the lowest average current result (0.276$\mu A$). The ratio of RFC-to-LOCO average currents closely matches their power consumption ratio despite similar transistor counts. These results demonstrate that the proposed OSC can indeed reduce short-circuit current and power in the LOCO latch.

### 4.3 PVT Variation Analysis

Process variation, voltage scaling, and temperature altering (PVT) can significantly impact the performance of modern circuits. To evaluate the robustness of the proposed LOCO latch under these conditions, a fluctuation range of the threshold voltage from 0.01 V to 0.08V, a supply voltage variation from 0.5 V to 1.0V, and a temperature variation from -40°C to 150°C were applied in SPICE simulation. The symbols at the top of Fig. 9 represent the same latch designs for all subfigures.

Fig. 9(a) and (b) show the power and delay fluctuations under different threshold voltages, respectively. The power and delay of the proposed LOCO latch remain stable with the rising of threshold voltage, and are lower than other designs. Therefore, the proposed LOCO latch has low sensitivity to variation of process.

Fig. 9(c) and (d) show the power and delay fluctuations under different temperatures, respectively. The temperature ranges from -40°C to 150°C with a step of 10 degrees, which is for harsh application environment. The power of the proposed LOCO latch increases slightly with the rising of the temperature, which is still lower than rest designs. The delay of the proposed LOCO latch remains stable, while RFEL, STAHL, HIDER, and the standard latch have much delay augment. The proposed LOCO latch has low sensitivity to variation of temperature.

Fig. 9(e) and (f) show the fluctuations of power and delay under different supply voltages, respectively. As expected, a higher supply voltage generally leads to a higher power consumption but lower delay. The power of all compared designs raises with the increasing of supply voltage, however, the proposed LOCO latch's power consumption is always the lowest and increases at a slower rate. Furthermore, its delay remains stable across the voltage range. Therefore, the variation of supply voltages has low impact on the power and delay of the proposed LOCO latch.

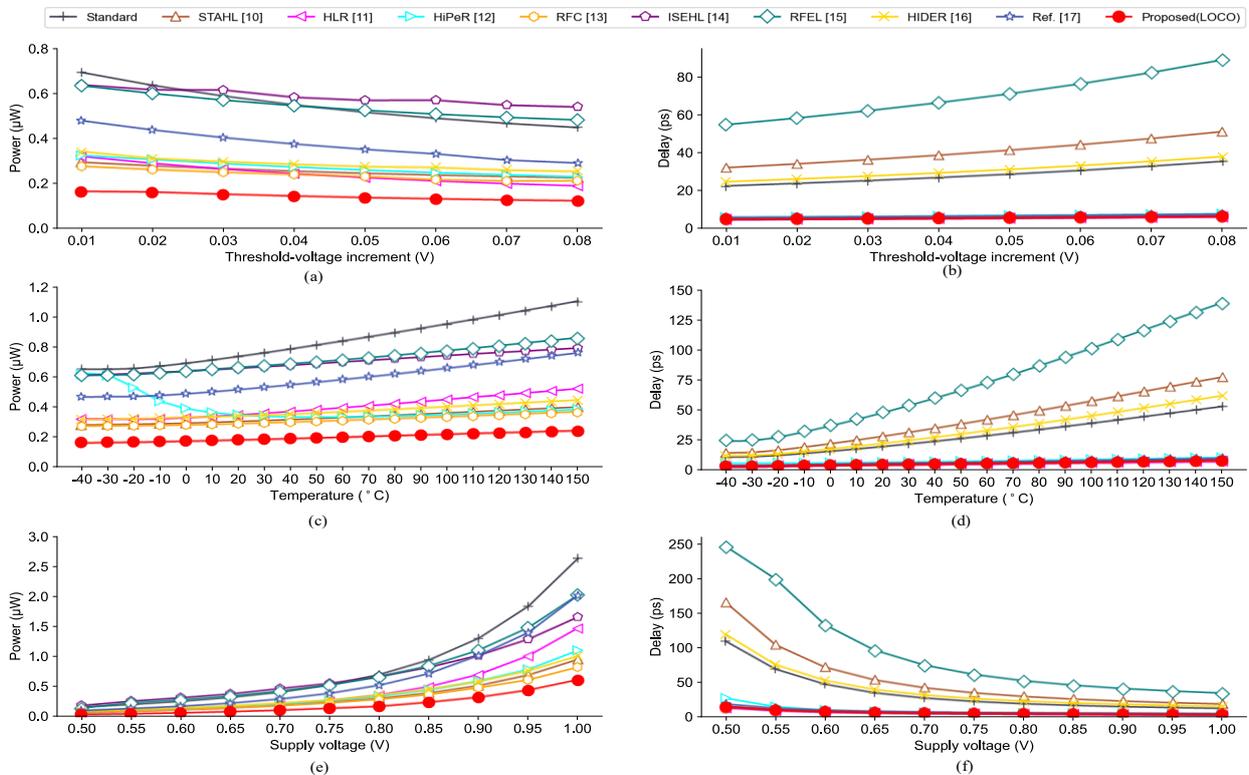

**Fig. 9** Impact of PVT variations on latch designs. Impact of various threshold voltages on **a** power and **b** delay, impact of various temperatures on **c** power and **d** delay, and impact of various voltages on **e** power and **f** delay

**Table 4** Standard deviation of the power and delay of latch designs

| Latch | Ref. | σ(PP) | σ(PD) | σ(TP) | σ(TD) | σ(VP) | σ(VD) |
|---|---|---|---|---|---|---|---|
| Standard | - | 7.89E-02 | 4.24 | 1.45E-01 | 13.27 | 0.75 | 28.41 |
| STAHL | [10] | 2.15E-02 | 6.18 | 3.74E-02 | 19.88 | 0.26 | 42.72 |
| HLR | [11] | 4.23E-02 | 0.40 | 6.59E-02 | **1.19** | 0.43 | **2.67** |
| HiPeR | [12] | 3.12E-02 | 0.55 | 8.49E-02 | 1.41 | 0.31 | 6.40 |
| Ref. [17] | [17] | 6.09E-02 | 0.55 | 9.50E-02 | 1.79 | 0.58 | 3.88 |
| RFC | [13] | 2.12E-02 | 0.47 | 2.89E-02 | 1.44 | 0.22 | 3.19 |
| ISEHL | [14] | 3.12E-02 | 0.47 | 5.70E-02 | 1.39 | 0.44 | 3.06 |
| RFEL | [15] | 4.89E-02 | 10.97 | 7.72E-02 | 36.57 | 0.56 | 67.76 |
| HIDER | [16] | 2.54E-02 | 4.17 | 3.94E-02 | 15.73 | 0.28 | 30.15 |
| LOCO | Proposed | **1.40E-02** | **0.42** | **2.33E-02** | 1.28 | **0.17** | 2.83 |

$$\sigma = \sqrt{\frac{\sum (X_i - \overline{X})^2}{N}} \quad (5)$$

The standard deviation (σ) is introduced to quantitatively measure the impact of PVT variations on the power and delay of all compared latch designs, which is shown in Eq. (5), where $N$ is the number of simulation cases, $X_i$ is the $i$-th simulation result, and $\overline{X}$ is an average value of $N$ cases. Table 4 shows these standard deviation results of each latch under different PVT variables. From left to right, "σ(PP)" is the standard deviation of power at different threshold voltages, "σ(PD)" is the standard deviation of delay at different threshold voltages, "σ(TP)" is the standard deviation of power consumption at different temperatures, "σ(TD)" is the standard deviation of delay at different temperatures, "σ(VP)" is the standard deviation of power at different supply voltages, and "σ(VD)" is the standard deviation of delay at different supply voltages. The proposed LOCO latch has the best "σ(PP)", "σ(PD)", "σ(TP)", and "σ(VP)" results. In the aspect of "$\sigma$(TD)", the proposed LOCO latch has the second-best result. In the aspect of "$\sigma$(VD)", the proposed LOCO latch has the third-best result. These quantitative deviation results strongly support the stability of the proposed LOCO latch under the impact of PVT variations.

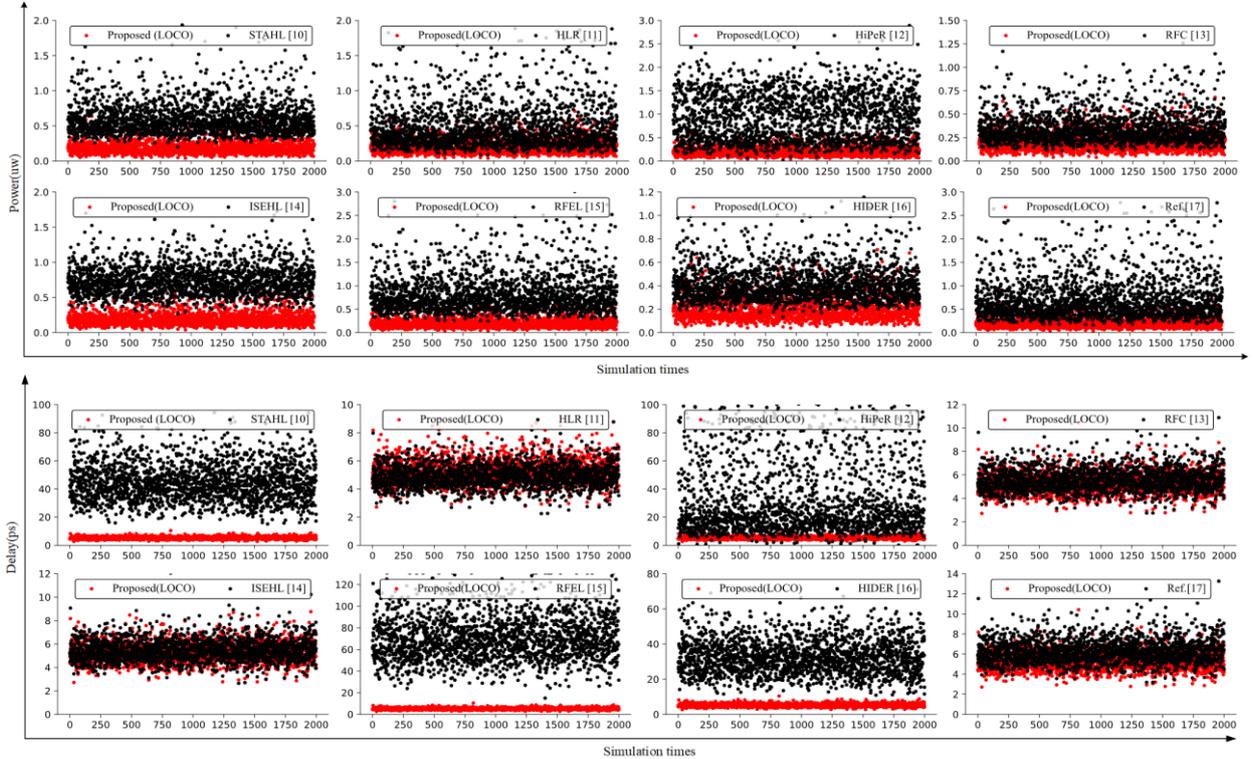

**Fig. 10** Monte Carlo simulation results on the power and delay

**Table 5** Standard deviation ($\sigma$) and Average deviation (*AD*)

| Latch | Ref. | Power | | Delay | |
|---|---|---|---|---|---|
| | | $\sigma$ | AD | $\sigma$ | AD |
| Standard | - | 0.47 | 0.35 | 52.42 | 13.27 |
| STAHL | [10] | 0.22 | 0.16 | 13.03 | 10.31 |
| HLR | [11] | 0.31 | 0.22 | **0.80** | **0.62** |
| HiPeR | [12] | 0.50 | 0.42 | 113.66 | 29.62 |
| Ref. [17] | [17] | 0.44 | 0.32 | 1.17 | 0.89 |
| RFC | [13] | 0.16 | 0.11 | 0.97 | 0.74 |
| ISEHL | [14] | 0.21 | 0.16 | 0.91 | 0.70 |
| RFEL | [15] | 0.39 | 0.29 | 19.98 | 15.57 |
| HIDER | [16] | 0.14 | 0.11 | 9.63 | 7.62 |
| LOCO | Proposed | **0.09** | **0.07** | 0.83 | 0.64 |

For further evaluating the impact of PVT variations on the proposed LOCO latch, two thousand times Monte Carlo simulations were performed with the threshold voltage scanned with a Gaussian distribution of ±10% at ±3$\sigma$ level, with the temperature scanned with a Gaussian distribution of ±3$\sigma$ level, and with the supply voltage scanned with a Gaussian distribution of ±20% at ±3$\sigma$ level. The 8 subfigures in Fig. 10 compare the distribution of the power results between the proposed LOCO latch and a hardened latch design, while the rest 8 subfigures compare their delay distribution results.

It can be seen from Fig. 10 that the power distribution of the proposed LOCO latch is relatively stable and at a low level. The delay fluctuation of the proposed LOCO latch is similar to HLR [11], RFC [13], ISEHL[14], and Ref. [17], and is better than others.

The standard deviation ($\sigma$) and the average deviation (*AD*) are used to quantitatively evaluate these Monte Carlo (MC) simulations, which are calculated by Eq. (5) and Eq. (6), respectively. Among them, *N* is the number of simulations, which is 2000, $X_i$ represents the *i*-th simulation value, and $\bar{X}$ is the average value. Table 5 shows the standard deviation and the average deviation results. A lower deviation result shows a lower sensitivity of PVT variations. The proposed LOCO latch has the lowest $\sigma$ and *AD* results in terms of power, which is the best among all compared designs. In terms of delay, the proposed LOCO latch is only second to HLR, which is a soft error tolerant hardened latch design.

$$AD = \frac{\sum |X_i - \bar{X}|}{N} \quad (6)$$

In summary, the proposed LOCO latch demonstrates superior stability, lower power consumption, and lower propagation delay across a wide range of PVT conditions, confirming its robustness for modern circuit applications.

## 5 Conclusions

This paper has proposed a novel Output-Split C-element (OSC) that protects both its input and output nodes from soft errors. The proposed low-cost SNU-self-resilient latch (LOCO) is the first design to employ OSCs for achieving soft error resilience with low overhead. By incorporating clock gating technology and high-speed transmission paths, the proposed LOCO design significantly reduces both power consumption and delay. Compared with state-of-the-art SNU-resilient hardened latch designs, the proposed LOCO latch demonstrates superior performance: it reduces area by 19.30%, power consumption by 63.58%, delay by 74.47%, and the power-delay-product (PDP) by 92.34% on average, respectively. Furthermore, the proposed LOCO latch exhibits better stability with respect to variations in PVT as well. Extensive PVT and MC simulations have proved the stability of the proposed LOCO latch under various PVT variations.

**Acknowledgments** This work was supported in part by KAKENHI under Grant 25K15042; in part by JPJSBP under Grant 120237413; in part by the NSFC-JSPS Exchange Program under Grant 62111540164; and in part by the JSPS Grant-in-Aid for Scientific Research (B) under Grant 21H03411.

**Data Availability** The datasets generated and analyzed during the current study are available from the corresponding author on reasonable request.

**Declarations**

**Conflicts of Interests** The authors declare that they have no known competing financial interests or personal relationships that could have appeared to influence the work reported in this paper.

## REFERENCES

1. Ebara M, Yamada K, Kojima K, Furuta J, Kobayashi K (2019) Process dependence of soft errors induced by alpha particles, heavy ions, and high energy neutrons on flip flops in FDSOI. IEEE J Electron Devices Soc 7:817–824. https://doi.org/10.1109/JEDS.2019.2907299
2. Lin D, Wen C (2020) DAD-FF: Hardening designs by delay-adjustable d-flip-flop for soft-error-rate reduction. IEEE Trans VLSI Syst 28(4):1030–1042. https://doi.org/10.1109/TVLSI.2019.2962080
3. Gadlage M, Roach A, Duncan A, Williams A, Bossev D, Kay M (2016) Soft errors induced by high-energy electrons. IEEE Trans Device Mater Reliab 17(1):157–162. https://doi.org/10.1109/TDMR.2016.2634626
4. Rajaei R (2016) Single event double node upset tolerance in mos/spintronic sequential and combinational logic circuits. Microelectron Reliab 69:109–114. https://doi.org/10.1016/j.microrel.2016.12.003
5. Yan A, Li Z, Huang Z, Ni T, Cui J, Girard P, Wen X (2024) MURLAV: a multiple-node-upset recovery l

atch and algoritm-based verification method. IEEE Trans Comput-Aided Des Integr Circuits Syst 43(7):2205–2214. https://doi.org/10.1109/TCAD.2024.3357593
6. Talpes E, Sarma D, Venkataramanan G, Bannon P, McGee B, Floering B, Jalote A, Hsiong C, Arora S, Gorti A, Sachdev G (2020) Compute solution for tesla's full self-driving computer. IEEE Micro 40(2):25–35. https://doi.org/10.1109/MM.2020.2975764
7. Iwashita H, Funatsu G, Sato H, Kamiyama M, Wender S, Pitcher E, Kiyanagi Y (2020) Energy-resolved soft-error rate measurements for 1–800 mev neutrons by the time-of-flight technique at lansc. IEEE Trans Nucl Sci 67 (11):2363–2369. https://doi.org/10.1109/TNS.2020.3025727
8. Lin S, Kim Y, Lombardi F (2010) Design and performance evaluation of radiation hardened latches for nanoscale cmos. IEEE Trans VLSI Syst 19(7):1315–1319. https://doi.org/10.1109/TVLSI.2010.2047954
9. Bisdounis L, Koufopavlou O (2002) Short-circuit energy dissipation modeling for submicrometer CMOS gates. IEEE Trans Circuits Syst I Fundam Theory Appl 47(9):1350-1361. https://doi.org/10.1109/81.883330
10. Ma R, Holst S, Wen X, Yan A, Xu H (2022) Evaluation and test of production defects in hardened latches. IEICE Trans Info and Syst E105D(5):996–1009. https://doi.org/10.1587/transinf.2021EDP7216
11. Nan H, Choi K (2011) High Performance, Low Cost, and Robust Soft Error Tolerant Latch Designs for Nanoscale CMOS Technology. IEEE Transactions on Circuits and Systems I: Regular Papers 59(7):1445-1457. https://doi.org/10.1109/TCSI.2011.2177135
12. Omana M, Rossi D, Metra C (2010) High-performance robust latches. IEEE Trans Comput 59(11):1455–1465. https://doi.org/10.1109/TC.2010.24
13. Yan A, Liang H, Huang Z, Jiang C, Yi M (2015) A self-recoverable, frequency aware and cost-effective robust latch design for nanoscale cmos technology. IEICE Trans Electron 98(12):1171–1178. https://doi.org/10.1587/transele.E98.C.1171
14. Liang H, Wang Z, Huang Z, Yan A (2014) Design of a radiation hardened latch for low-power circuits. In: Proc. 23rd IEEE Asian Test Symposium, pp. 19–24. https://doi.org/10.1109/ATS.2014.16
15. Yan A, Liang H, Huang Z, Jiang C, Ouyang Y, Li X (2016) An seu resilient, set filterable and cost effective latch in presence of pvt variations. Microelectron Reliab 63:239–250. https://doi.org/10.1016/j.microrel.2016.06.004
16. Ma R, Holst S, Xu H, Wen X, Wang S, Li J, Yan A (2025) Highly Defect Detectable and SEU-Resilient Robust Scan-Test-Aware Latch Design. IEEE Trans VLSI Syst 33(2):449-461. https://doi.org/10.1109/TVLSI.2024.3467089
17. Kumar S, Mukherjee A (2021) A self-healing, high performance and low-cost radiation hardened latch design. In: Proc. 2021 IEEE International Symposium on Defect and Fault Tolerance in VLSI and Nanotechnology Systems (DFT), pp. 1–6. https://doi.org/10.1109/DFT52944.2021.9568359
18. Messenger G (1982) Collection of charge on junction nodes from ion tracks. IEEE Trans Nucl Sci 29(6):2024–2031. https://doi.org/10.1109/TNS.1982.4336490
19. Jiang J, Xu Y, Ren J, Zhu W, Lin D, Xiao J, Kong W, Zou S (2018) Low-cost single event double-upset tolerant latch design. Electronics letters 54(9):554–556. https://doi.org/10.1049/el.2018.0558
20. Yan A, Lai C, Zhang Y, Cui J, Huang Z, Song J, Wen X (2018) Novel low cost, double-and-triple-node-upset-tolerant latch designs for nano-scale CMOS. IEEE Trans Emerg Top Comput 9(1):520-533. https://doi.org/10.1109/TETC.2018.2871861
21. Andjelkovic M, Krstic M, Kraemer R, Veeravalli VS, Steininger A (2017) A critical charge model for estimating the SET and SEU sensitivity: a muller C-Element case study. In: Proc. 2017 IEEE Asian Test Symp (ATS), pp. 82–87. https://doi.org/10.1109/ATS.2017.27
22. Watkins A, Tragoudas S (2017) Radiation hardened latch designs for double and triple node upsets. IEEE Trans Emerging Top Comput 8(3):616–626. https://doi.org/10.1109/TETC.2017.2776285
23. Zhao W, Cao Y (2007) Predictive technology model for nano-cmos design exploration. In: Proc. 2006 1st International Conference on Nano-Networks and Workshops, pp. 1–5. https://doi.org/10.1145/1229175.1229176
24. Huang Z, Wang H, Ang Y, Liang H, Ouyang Y, Ni T (2021) A high-speed and triple-node-upset recovery latch with heterogeneous interconnection. Microelectron J 118:105290. https://doi.org/10.1016/j.mejo.2021.105290
25. Predictive Technology Model for Spice [Online]. Available: https://web.archive.org/web/20100420181637/http://ptm.asu.edu/